\documentclass{ws-procs975x65}
\usepackage{tikz-cd}

\def\beq{\begin{equation}}
\def\eeq{\end{equation}}

\begin{document}
\title{\Large Cosmological Principle in Newtonian Dynamics}
\author{David Benisty}
\address{Physics Department, Ben-Gurion University of the Negev, Beer-Sheva 84105, Israel\\
Frankfurt Institute for Advanced Studies (FIAS), Ruth-Moufang-Strasse 1, 60438 Frankfurt am Main, Germany\\
E-mail: benidav@post.bgu.ac.il}
\author{Eduardo I. Guendelman} 

\address{
Physics Department, Ben-Gurion University of the Negev, Beer-Sheva 84105, Israel
\\
Frankfurt Institute for Advanced Studies (FIAS), Ruth-Moufang-Strasse 1, 60438 Frankfurt am Main, Germany,\\
Bahamas Advanced Study Institute and Conferences, 4A Ocean Heights, Hill View Circle, Stella Maris, Long Island, The Bahamas\\
E-mail: guendel@bgu.ac.il}

\begin{abstract}
A correspondence between the Equivalence principle and the homogeneity of the universe is discussed. In Newtonian gravity, translation of co-moving coordinates in a uniformly expanding universe defines an accelerated frame. A consistency condition for the invariance of this transformation which requires a well defined transformation for the Newtonian potential, yields the Friedmann equations. All these symmetries are lost when we modify NSL (Newton's Second Law) or the Poisson equation. For example by replacing NSL with non-linear function of the acceleration the concept of relative acceleration is lost and the homogeneity of the universe breaks.
\end{abstract}


\bodymatter
\section{Introduction}
Newtonian gravity is a simple framework for studying cosmology \cite{Bondi,Kopeikin:2014qna,Kopeikin:2013kpa,Bergshoeff:2014uea}. Our analysis exploits the symmetries of Newton's equations which can be the basis for the cosmological principle. One well known symmetry is the "Galilean invariance" which corresponds to a transformation to other frame moving with respect to the original frame with constant velocity. Since the acceleration is invariant under this transformation, this symmetry does not imply any restriction on possible nonlinear generalization of Newton's equations on the acceleration dependence.

Here we show that more general symmetries that connect two different frames with relative acceleration are more useful. In particular, when one frame introduces a homogeneous expanding universe, where Hubble’s law is valid with respect to any point in the universe, we show that
the homogeneity of the universe emerges from a basic
symmetry of NSL which allows
us to introduce a uniform acceleration in space, but not
constant in time. In order to obtain a symmetry of the Newtonian equations of motion, the Newtonian potential is
also transformed accordingly as we go to the new accelerated frame.

\section{Basic Derivation}
Newtonian dynamics can derive the cosmological Friedmann equation. Therefore we emphasize here the simplest point of view which starts by taking into account the expansion of the universe from the Hubble’s law:
\begin{equation} \label{expansion}
\frac{dR}{dt}=H R,
\end{equation}
where $H$ is the Hubble parameter and $R$ is the scale parameter of the universe. For a spherically symmetric object with a radius $r$ with a test mass $m$ outside the sphere, the total energy reads:
\begin{equation} \label{energy}
E=\frac{1}{2}m r^2 \left(\frac{dR}{dt} \right)^{2}-\frac{G M m}{r R}.
\end{equation}
Notice that we multiply the radius $r$ by the scale factor $R$. Inside the sphere the mass is being:
\begin{equation}
M = \frac{4}{3}\pi r^3 R^{3}\rho,   
\end{equation}
where $\rho$ is the density inside the sphere. Eq. (\ref{energy}) takes the form:
\begin{equation} \label{H1}
H^{2}=\frac{8\pi G}{3}\rho - \frac{k^{2}}{R^{2}}
\end{equation}
where $k \equiv 2E/ (m r^2) = \text{Const}$. 
This is the first Friedmann equation which does not depend on the size of the sphere $r$. $E$ is the total conserved energy of the test particle. However, $k$ is the energy per $\frac{1}{2}  m r^2$ that alleviates the dependence on the chosen size of the sphere.

To get the second Friedmann equation we use the conservation of the energy in an expanding volume $V$. The pressure does work equal to $p dV$ which decreases the energy in $V$ by that amount. The conservation gives:
\begin{equation} \label{work}
d\left( \rho \frac{4}{3}\pi R^{3}\right)=-p\,d\left(\frac{4}{3}\pi R^{3} \right).
\end{equation}
which is the continuity equation. By taking a derivative of Eq. (\ref{H1}) and replacing $R \dot\rho$ from Eq. (\ref{work}), we obtain:
\begin{equation} \label{second}
\frac{d^{2}R}{dt^{2}}=-\frac{4\pi G}{3}(\rho+3p)R
\end{equation}
From our point of view the crucial ingredient is how the Newtonian potential enters the derivation. But we can obtain the same equation by moving to an accelerated frame.
\section{Newtonian Potential transformation to an accelerated frame}
Newton's equations are invariant under Galilean transformation which relates two systems through a uniform velocity. Because the acceleration is invariant under Galilean transformation, it does not impose any constraint on possible non linear dependence on acceleration for generalizations of NSL. In addition to the Galilean transformation, Newtonian theory allows to transform for a uniformly accelerated frame. Indeed, as observed in \cite{Shucking,Arka} the NSL with the law of gravitation: 
\begin{equation}
\ddot{X}_{i}+\frac{\partial}{\partial X_{i}}\Phi(X_{j},t)\,=\,0
\end{equation}
holds invariant under a transformation into an accelerated coordinate frame:
\begin{equation}
\ddot{X}_{i}+\frac{\partial}{\partial X_{i}}\Phi(X_{j},t)=0=\ddot{X}'_{i}+\frac{\partial}{\partial X'_{i}}\Phi'(X'_{j},t),
\end{equation}
while transformation we introduce an arbitrary uniform acceleration, in the following way:
\begin{equation}\label{aceTra}
\ddot{X}_{i}'(t)-\ddot{X_{i}}(t)=g_{i}(t)
\end{equation}
\begin{equation}\label{potTra}
\Phi'(X_{j}',t)=\Phi[X_{i}(x_{j}'),t]-g_{k}(t)X_{k}(x_{j}')+h(t)
\end{equation}
where $\forall i,j,k=1,2,3.$. $h(t)$ is a constant of integration with respect to the space gradient, but may depend on time.
\begin{figure}[t!]
 	\centering
\[
\begin{tikzcd}
Global\,Equivalence\,Principle \arrow[dr,swap,""] && Poisson\, equation \arrow[dl,""]  \\
& Homogeneity\,of\, the\,universe
\end{tikzcd}
\]
\caption{The global Equivalence Principle with Poisson equation yields the homogeneity of the universe.}
 	\label{fig}
 \end{figure}
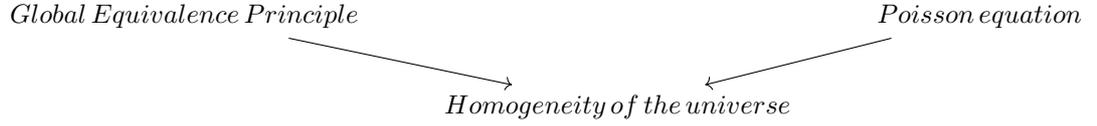
The Poisson equation is also invariant under (\ref{potTra}):
\begin{equation}\label{Poisson}
\Delta \Phi=4\pi G\rho (t).
\end{equation}
In the global equivalence principle a transformation into a frame with a relative spatially uniform acceleration, equivalents to introduce a uniform gravitational field. This can be compensated by a linear contribution to the Newtonian potential which generates a gravitational field with the opposite sign (\ref{potTra}). The second crucial element is the isotropic solution for the Poisson equation (\ref{Poisson}) that reads: 
\begin{equation}\label{tranPoten}
\Phi=\frac{2\pi G\rho (t)}{3}X_{i}X_{i}.
\end{equation}
Due to the uniform background, the coordinates may be any coordinate system. However, it is simpler to consider the Cartesian coordinate system.
Notice that our assumption that the density $\rho$ is a function of time comes from the cosmological principle. Eq. (\ref{tranPoten}) doesn't single out a special point in the universe, since the transformations (\ref{aceTra})-(\ref{potTra}) imply for cosmology that all points are on an equal footings. Therefore the potential is well defined for any arbitrary origin.

The variable $x_i$ is defined from assuming a uniform expansion of the volume:
\begin{equation}
\label{comExp}
 X_i = x_i R(t)   
\end{equation}
where $R(t)$ is the time dependent scale factor of the universe. Now we check the validity of the translation invariance of the space:
\begin{equation}\label{trans}
x_i \rightarrow x_i + c_i,
\end{equation}
where $c_i $ denotes a constant vector. Therefore the position expands by the relation: 
\begin{equation}
X_i \rightarrow X_i + c_i R(t) = X'_i,
\end{equation}
and the local acceleration transforms as:
\begin{equation}
\ddot{X}_i \rightarrow \ddot{X}_i + c_i \ddot{R}(t).
\end{equation}
From Eq. (\ref{aceTra}) we identify the relative acceleration $g_i(t)$ as:
\begin{equation}\label{gi}
g_i(t) = c_i \ddot{R}(t).
\end{equation}
The transformation of the gravitational potential, from Eq. (\ref{tranPoten}), reads:
\begin{equation}\label{tranPotenS}
\begin{split}
\Phi'(X')=\frac{2\pi G\rho}{3}X'_{i}X'_{i} \\= \frac{2\pi G\rho}{3} (X_i + c_i R(t))(X_i + c_i R(t)) \\ = \frac{2\pi G\rho}{3} X_i X_i + \frac{4\pi G\rho}{3} X_i c_i R(t) +  \frac{2\pi G\rho}{3}c_i c_i R(t)^2.   
\end{split}
\end{equation}
This transformation corresponds to Eq. (\ref{potTra}) only for:
\begin{equation}\label{gii}
g_i(t) = -\frac{4\pi G\rho}{3} c_i R(t) , \quad h(t) = \frac{2\pi G\rho}{3}c_i c_i  
\end{equation}
For a consistency between Eq. (\ref{gi}) and Eq. (\ref{gii}) we get:
\begin{equation}
c_i \ddot{R}(t)=-\frac{4\pi G\rho}{3} c_i R(t),
\end{equation}
which reduces to the relation:
\begin{equation}
\frac{\ddot{R}(t)}{R(t)} = -\frac{4\pi G \rho}{3}.
\end{equation}
This relation is the known Friedmann equation. For a transformation into the accelerated frame, the extra terms from Eq. (\ref{potTra}) can be eliminated by the transformation (\ref{tranPoten}). Similar to the equivalence principle that produces a potential from a relative acceleration, here we eliminate the potential by shift into the accelerated frame. This proves that the potential is good for any arbitrary origin. Since we started with isotropy with respect to a particular point and we got homogeneity of the Universe, we obtain homogeneity and isotropy for each point in the Universe. This is the Newtonian analog for the homogeneity and isotropy of the FRW spacetime.

\section{Modified Newtonian Dynamics}
MOND is an alternative successful explanation to the flat rotation curves for galaxies. By violating NSL at low accelerations the Tully-Fisher relation is recovered, without introducing additional dark matter \cite{Tully:1977fu,McGaugh:2000sr}. By introducing a non-linear dependence on the accelerations, as MOND assumes, the homogeneity of the universe breaks and a uniform perfect fluid is impossible. 

The fundamental approach of MOND \cite{Milgrom:1983zz,Milgrom:1983ca,Sanders:2002pf,Felten} consists of changing Newton's Second Law (NSL) instead of adding dark matter, in order to explain the flat rotation curve of galaxies. Millgrom's suggestion was to modify NSL by considering a function of the acceleration:
\begin{equation}
F = m a \, \mu \left(\frac{a_0}{a}\right)
\end{equation}
The $\text{simple}$ MOND representation uses the function:
\begin{equation}
\mu \left(\frac{a_0}{a}\right) = \left[ 1+ \frac{a_0}{a} \right]^{-1},
\end{equation}
while the $\text{standard}$ representation uses the function:
\begin{equation}
\mu \left(\frac{a_0}{a}\right) = \left[ 1+ \right(\frac{a_0}{a}\left)^2 \right]^{-1/2}.
\end{equation}
For $a \gg a_0$ both functions $\mu \left(\frac{a}{a_0}\right) \rightarrow 1$, which reproduces the NSL. The \text{deep}-MOND regime reads the limit $a \ll a_0$. In the deep-MOND regime both functions are reduced to $\mu \left(\frac{a}{a_0}\right) \rightarrow \frac{a}{a_0}$, which yields the galactic flat rotation curve, with modifies NSL:
\begin{equation}
F = m \left(\frac{a^2}{a_0}\right)
\end{equation}

Another approach of modification is to keep the Newton's equation, but to change the Poisson equation (\ref{Poisson}) with the the function $\mu(\frac{\bigtriangledown \Phi}{a_0})$, as \cite{Bekenstein:1984tv}. The modified Poisson equation reads:
\begin{equation}\label{modPoi}
\bigtriangledown \cdot \left(\mu(\frac{|\bigtriangledown \Phi|}{a_0}) \bigtriangledown \Phi \right) = 4 \pi G \rho.
\end{equation}
In this case the transformation (\ref{aceTra}) - (\ref{potTra}) satisfies the symmetry of NSL, but the modified Poisson equation is not invariant under the part of the symmetry that involves the transformation of the Newtonian potential (\ref{potTra}). Indeed, when we solve the modified Poisson equation (\ref{modPoi}) for $\rho = \rho(t)$ we don't get a quadratic form of the potential as in Eq. (\ref{tranPoten}). Then, when we attempt the solution with homogeneous expansion of some co-moving coordinates as in (\ref{comExp}) and consider the translation transformation (\ref{trans}), we see that there is no way to make the transformation of the modified gravitational potential in accordance with the form of (\ref{potTra}). Therefore the homogeneity of the universe is violated.

Possible generalizations of MOND allow homogeneous and isotropic cosmologies. \cite{Milgrom:2009gv,Clifton:2010xv,Deffayet:2014lba} discuss the relativistic version of MOND which in general are  differently formulated. When the theory begins from a covariant action principle, the cosmological solution is possible with TeVeS \cite{Bekenstein:2004ne,Chaichian:2014qba} and MoG \cite{Moffat:2011ya,Moffat:2011rp,Moffat:2015bda} and other alternatives \cite{Zlosnik:2017xpr,Jamali:2018uij,Chaichian:2014dfa,Benisty:2018qed}. 

Although cosmology is maybe well defined in the relativistic versions of MOND, for practical applications these theories are not used in phenomenological applications. For example, \cite{Banik:2018zhm} studies the effect of the cosmological constant for the Local Group of galaxies, in the framework on MOND. \cite{Banik:2018zhm} imposes a cosmological version of MOND, which is not based on any relativistic version of MOND. On one hand, \cite{Banik:2018zhm} introduces homogeneous expansion of the background, using the scale parameter $a(t)$ function, implicitly assuming homogeneity of the universe. On the other hand, we have shown that the homogeneity of universe is inconsistent with MOND. \cite{Benisty:2019fzt} studies also the the effect of the cosmological constant in the Local Group of galaxies. However, when \cite{Benisty:2019fzt} studies the implication of MOND, they not invoke any cosmological effect using the scale factor, but \cite{Benisty:2019fzt} solves a spherically symmetric problem for a theory with a cosmological constant background. The contribution for the Newtonian limit yields a linear term $\sim \Lambda r$. From our analysis, the approach of \cite{Benisty:2019fzt} is more consistent.

Previous studies connected MOND with cosmology as \cite{Sanders:1997we}. One from the fundamental connections was the relation between the critical acceleration of MOND and the observed value of the Hubble constant:
\begin{equation}
a_0 \sim H_0 \, c.
\end{equation}
However, as we will see there is a deeper contradiction between the MOND approach and the approach of cosmology. A review on the problems for MONDian cosmology is in \cite{Sanders:2005pi}. \cite{Scott:2001td} mentions some cosmological difficulties with MOND and claims that MOND may violate the Cosmological Principle. Here we show that MOND or any modification to NSL or the Poisson equation violates directly the cosmological principle explicitly from a violation of symmetries that are satisfied in the Newtonian case and that guarantee the existence of homogeneous Universes, but that that are absent in the MOND case.

\section{Epilogue}
Here we show the correspondence between the Equivalence principle and the homogeneity of the universe. The Equivalence principle here is understood 
as the invariance of Newton's equation under the introduction of a global, uniform acceleration all over space. We examine the application of these transformations in Newtonian Cosmology and see that they appear naturally when we consider translation of "co-moving coordinates". We show then that under the Newtonian gravity, translation of co-moving coordinates in a uniformly expanding universe defines a new accelerated frame. Using the simple quadratic coordinates term in the solution of the standard Poisson equation for a Universe with constant density, the condition for the invariance of this transformation yields the second Friedmann equation. This implies that the cosmological principle can be satisfied with NSL.

All these symmetries get lost when we modify NSL and/ or the Poisson equation. By replacing NSL with non linear function of the acceleration, as Modified Newtonian Dynamics suggested, the concept of relative acceleration is lost and the symmetry (\ref{aceTra}-\ref{potTra}). As a consequence the homogeneity of the universe is impossible.

Conceptually, relative acceleration in MOND does not exist. Since we would like to approach for any galaxy in the limit of small accelerations (for halo of the galaxies), we assume there is a different behavior between from large and small accelerations. However, from the basis of the cosmological principle there is no difference between accelerated and inertial frames, or between small and large accelerations. Therefore, MOND is not a complete theory that should be amended to preserve the cosmological principle. However, a local version of MOND could be use as a good toy model, but not describing a uniform universe.

Small violations of the equivalence principle also occur for quantum effects in curved spacetime. \cite{Singleton:2011vh,Singleton:2016yal} show that one can get violations of the equivalence principle by comparing the Hawking radiation between uniformly accelerated frame and a gravitational field, while the violation goes away in the horizon. This is similar to the violation of the equivalence principle in the present paper as one goes between the limit $a \gg a_0$ which is consistent with the equivalence principle, and the limit $a \ll a_0$ where MOND is in play and the equivalence principle with the homogeneity is violated.

\section*{Acknowledgments}
This article is supported by COST Action CA15117 "Cosmology and Astrophysics Network for Theoretical Advances and Training Action" (CANTATA) of the COST (European Cooperation in Science and Technology). D.B.\ and E.G. \ are partially supported by COST Actions CA16104 and CA18108. We are grateful to M. Millgrom and M. Chaichian and to the referees for helpful and deep discussions.

\end{document}